\documentclass[10pt,letterpaper]{article}

\usepackage{cogsci}
\usepackage{graphicx}
\cogscifinalcopy % Uncomment this line for the final submission 

\usepackage{pslatex}
\usepackage{apacite}
\usepackage{booktabs}
\usepackage{float} % Roger Levy added this and changed figure/table
\title{Towards Cognitive Synergy in LLM-Based Multi-Agent Systems:\\Integrating Theory of Mind and Critical Evaluation}
 
\author{{\large \bf Adam Kostka (adam.kostka.stud@pw.edu.pl)} \\
  Faculty of Electronics and Information Technology \\
  Warsaw University of Technology
  \AND {\large \bf Jarosław A. Chudziak (jaroslaw.chudziak@pw.edu.pl)} \\
  Faculty of Electronics and Information Technology \\
  Warsaw University of Technology}

\begin{document}

\maketitle

\begin{abstract}
Recently, the field of Multi-Agent Systems (MAS) has gained popularity as researchers are trying to develop artificial intelligence capable of efficient collective reasoning. Agents based on Large Language Models (LLMs) perform well in isolated tasks, yet struggle with higher-order cognition required for adaptive collaboration. Human teams achieve synergy not only through knowledge sharing, but also through recursive reasoning, structured critique, and the ability to infer others’ mental states. Current artificial systems lack these essential mechanisms, limiting their ability to engage in sophisticated collective reasoning. This work explores cognitive processes that enable effective collaboration, focusing on adaptive theory of mind (ToM) and systematic critical evaluation. We investigate three key questions. First, how does the ability to model others’ perspectives enhance coordination and reduce redundant reasoning? Second, to what extent does structured critique improve reasoning quality by identifying logical gaps and mitigating biases? Third, the interplay of these mechanisms can lead to emergent cognitive synergy, where the collective intelligence of the system exceeds the sum of its parts. Through an empirical case study on complex decision making, we show that the integration of these cognitive mechanisms leads to more coherent, adaptive, and rigorous agent interactions. This article contributes to the field of cognitive science and AI research by presenting a structured framework that emulates human-like collaborative reasoning MAS. It highlights the significance of dynamic ToM and critical evaluation in advancing multi-agent systems’ ability to tackle complex, real-world challenges.

\end{abstract}

\section{Introduction}

Rapid progress in large language models (LLMs) has allowed human-level performance in some activities of natural language, including translation, summarization, and problem-solving~\cite{naveed2024comprehensiveoverviewlargelanguage}. However, LLMs continue to face challenges in multi-agent collaborative environments involving effective coordination, rigorous evaluation, and adaptive reasoning~\cite{qiu2024collaborativeintelligencepropagatingintentions}. Unlike human teams, which achieve cognitive synergy by integrating diverse knowledge, iterative critique, and dynamic mutual understanding (Theory of Mind), LLM-based systems often lack mechanisms for structured, adaptive coordination, risking fragmented solutions.

\begin{figure}[t]
\centering
\includegraphics[width=0.85\linewidth]{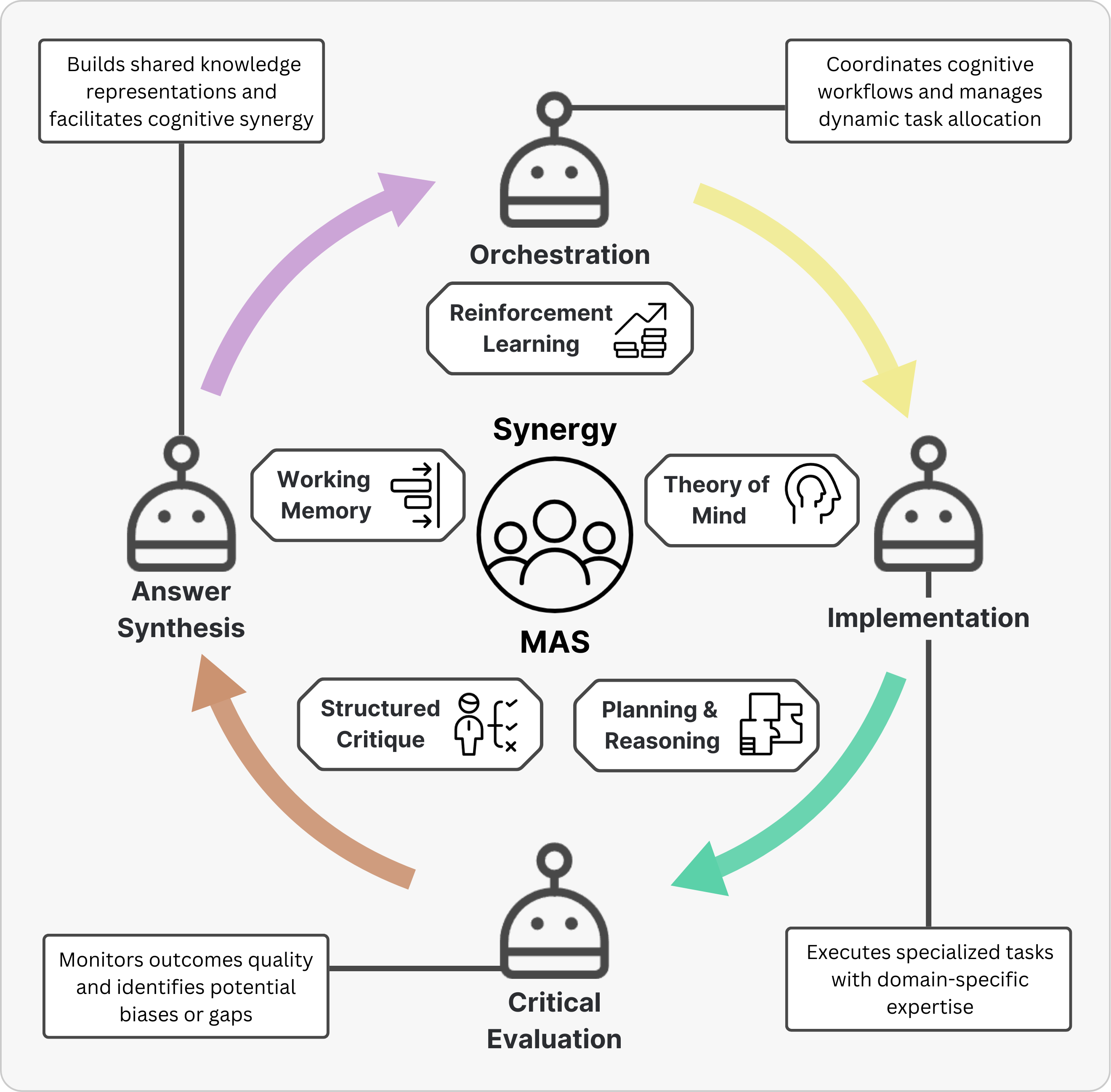}
\caption{Conceptual Cycle of Cognitive Synergy in Multi-Agent Systems}
\label{fig:intro}
\end{figure}

Theory of Mind (ToM) and structured critique are two significant mechanisms that can enable cognitive synergy in human collaboration. ToM allows individuals to infer others' beliefs, intentions, and knowledge states~\cite{PremackWoodruff1978, Wellman1990}, facilitating effective coordination and redundancy reduction~\cite{dech2010}. Structured critique, within this context, refers to a systematic process where designated agents or internal mechanisms evaluate the reasoning, evidence, and potential biases within the arguments presented by other agents, aiming to improve rigor and identify logical weaknesses, analogous to processes like peer review and argumentation in human reasoning~\cite{MercierSperber2011, MERCIER2016689}. Despite playing a central role in human cooperative endeavors, such mechanisms are hugely underrepresented in today's multiagent AI systems, thus limiting their ability for sophisticated cooperative reasoning.

Most modern LLM-based Multi-Agent Systems (MAS) also tend to rely on static role assignments and task divisions, not representing adaptively the shifting perspectives of their counterparts. Recent approaches that incorporate LLMs have improved interactions based on language~\cite{xi2023rise}, yet, such approaches frequently fail to achieve genuine coordination, since agents do not infer each other's reasoning states or recursively refine their arguments through systematic criticism~\cite{li2024challengesfacedlargelanguage}.

This work presents a MAS framework implementing cognitive synergy via dynamic ToM and structured critique. Figure~\ref{fig:intro} provides a conceptual overview of the interacting cognitive functions that can lead to synergy in MAS. Our specific approach features expert agents, a Critic, an Orchestrator, and an Integrator using a knowledge base and Clingo solver. The goal is synergy through integrated contributions, critique, and adaptive coordination.

A case study on strategic business decision making investigates: 1) How dynamic ToM affects coordination and redundancy. 2) How a Critic Agent improves reasoning quality and bias mitigation. 3) Whether these mechanisms yield emergent cognitive synergy, exceeding the sum of parts.

This work advances MAS design by combining cognitive science principles with AI concepts to present a framework intended to instantiate key aspects of collaborative human reasoning. The results underscore specialization, iterative critique, and adaptive coordination in AI systems for sophisticated decision-making.

\section{Related Work}

Collaborative AI system architectures have long been motivated by human cognition, and MAS research has long focused on task decomposition and pre-coordinated strategies to coordination~\cite{Wooldridge_2009,jennings1998roadmap}. Although such systems have been effective for structured workflows, they do not meet the flexible and adaptive collaboration of human teams. More recent attempts to integrate LLMs into MAS frameworks have improved language comprehension and communication~\cite{Li2024,muthusamy-etal-2023-towards}, yet much work remains to be done to improve collaborative intelligence.

% start
One of the most serious limitations of current LLM-Based Multi-Agent Systems is the poor development of Theory of Mind (ToM), the ability to represent such things as knowledge states, intentions, and cognitive processes of other agents~\cite{PremackWoodruff1978, Wellman1990}. In human collaboration, ToM often helps parties anticipate others' contributions, align views, and avoid duplicating effort. Developmental psychology has shown that ToM is crucial in collaboration and social learning~\cite{tom2005}. In multi-agent systems, dynamic ToM could enable iteratively improved responses, improving coordination and efficiency between agents~\cite{synergymas}.

Another part of fruitful collaboration is critical evaluation. Structured critique brings much improvement to human decision making, such as through peer review and deliberative reasoning processes~\cite{MERCIER2016689}. Most existing designs of MAS do not have such formally implemented mechanisms of critique, which could help reduce fallacious reasoning and weak conclusions. Some new frameworks offer self-correction for LLM-based systems~\cite{pan-etal-2023-logic}, but very few include an autonomous critic agent to perform structured critique and systematically assess the robustness of reasoning.

Mixture-of-Agents (MoA) models~\cite{wang2024mixtureofagentsenhanceslargelanguage} compose agents using large language models in a hierarchy, where each agent improves its responses by learning from past interactions. This framework supports dynamic answers, and also helps achieve better performance on very complex reasoning tasks. MoA mostly enhances the overall reasoning ability of different agents and focuses less on inter-agent evaluation or Theory of Mind~\cite{li-etal-2023-theory}.

Sparse Mixture-of-Agents (SMoA)~\cite{li2024smoaimprovingmultiagentlarge} builds on MoA. SMoA optimizes efficiency via selective communication methods (like Response Selection, Early Stopping) to minimize interactions between agents, thus reducing costs while preserving view diversity. Like MoA, however, SMoA primarily centers on optimizing agents and workload allocation, offering no deeper integration of structured criticism or ToM within these systems~\cite{DBLP:conf/iclr/GouSGSYDC24, cross2024hypothetical}. Higher-level cognitive mechanisms could, therefore, further enhance collaborative AI reasoning.

% end

\begin{figure}[t]
\centering
\includegraphics[width=1\linewidth]{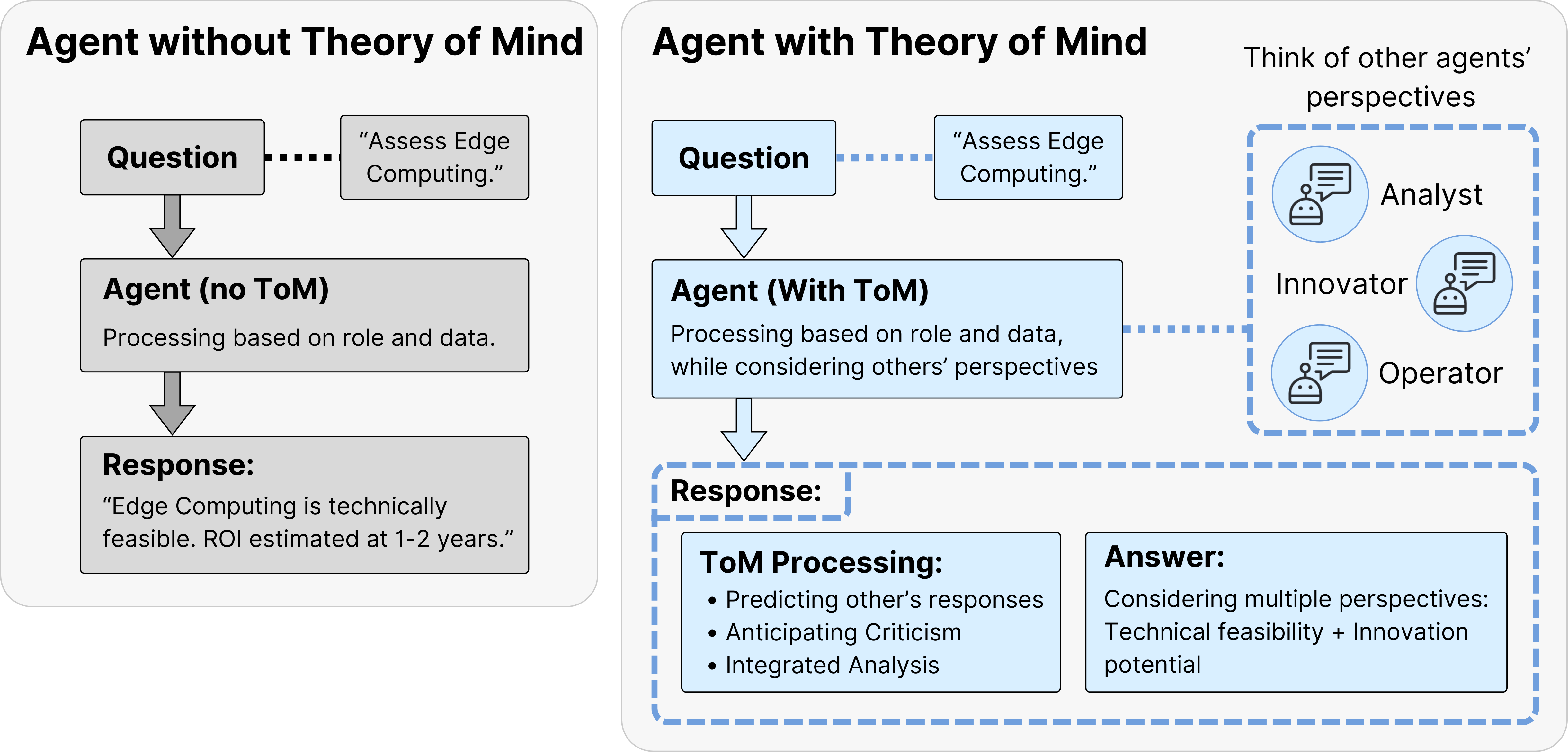}
\caption{Illustrative Comparison of Hypothesized Agent Response Patterns With and Without Theory of Mind (ToM) Capabilities.}
\label{fig:tom}
\end{figure}

Our work extends these traditions by formally including cognitive mechanisms in the design of MAS. In contrast to previous work that assumed agent interactions to be static or task-specific, or focused on supporting human collaboration in specific domains like knowledge discovery through structured knowledge representation like ontologies~\cite{Choiski2009OntologicalLA}, our design includes dynamic ToM to track changing agent beliefs and a Critic Agent specifically designed to offer structured, sequential criticism. Combining these mechanisms with structured knowledge retrieval and formal reasoning, we outline a system that better approximates human collaborative inference, trading off specialization, criticism, and flexibility in coordination~\cite{paclic_elliottagents}.

\section{Methodology and Proposed Team Structure}

This research explores how cognitive processes, specifically ToM and structured criticism, can be integrated into multi-agent systems to promote collaborative reasoning. The model will evaluate the extent to which peer-view simulation and systematic inclusion of critical evaluations deepen reasoning, enhance coordination efficiency, and reinforce the robustness of decision-making. To study these effects, the system is placed in a strategic investment choice scenario where a mid-size tech company must decide how to allocate its research and development budget among three emerging technologies: Edge Computing, Quantum Computing, and Blockchain. This exercise balances technical feasibility, market opportunity, and commercial viability, thus creating a formal setting where agents must iteratively synthesize multiple perspectives, analyze trade-offs, and refine their proposals. The system is implemented in Python, using the LangGraph library to orchestrate agent interactions in a state machine, LangChain components for LLM interaction, Neo4j for graph-based knowledge storage, and Clingo for logical reasoning. Each agent is powered by a large language model using gpt-4o for our case study.

\subsection{Cognitive Mechanisms for Adaptive Collaboration}

The model operationalizes ToM through structured prompting. When enabled, agents' prompts instruct them to 1) briefly anticipate expected peer arguments based on known roles (e.g., the Realist focusing on costs) and 2) frame their own contribution to complement or counter these anticipated views, thereby enhancing integration and reducing redundancy (see Figure \ref{fig:tom}). This methodology simulates the human act of mentalizing, in which humans forecast topics and areas of knowledge that their collaborators likely have rather than modeling their internal states directly. Agents also need to consider how to make their own responses complement those of their teams to encourage collaborative reasoning.

To operationalize ToM, prompts explicitly instructed agents to first state their anticipation of a specific peer's likely contribution angle based on their designated role, before formulating their own response (e.g., 'Anticipating the Realist will focus on budget constraints, I argue...'). Similarly, the Critic Agent's prompt guided it to specifically check for logical consistency between turns, identify claims lacking explicit support within the dialogue, and flag overlooked constraints mentioned in the initial scenario.

When ToM and the Critic Agent are enabled, the agents also predict possible counterarguments. This mechanism encourages an anticipatory defense of claims, very much in line with human deliberation, where arguments are preemptively supported, expecting peer appraisal. The Critic Agent mimics a System 2 process, allowing for a slow deliberative evaluation of reasoning parameters for rational flaws, unsupported claims, and overlooked hazards, correcting for the biases commonly associated with System 1 intuition. While the Critic does not provide new knowledge, it criticizes tacit assumptions, causing agents to justify their arguments against those assumptions. This reflects the cognitive processes of collaborative human problem-solving, where criticism leads to better analyses and error correction.

The Integrator and Orchestrator agents govern inter-agent communication. The implementation of ToM through prompting allows agents to modify their reasoning dynamically and not stick to a fixed decision. The prompting-based methodology ensures adaptive flexibility as the conversational context changes, similar to how human teams modify reasoning strategies following peer feedback.

\begin{figure}[t]
\centering
\includegraphics[width=0.9\linewidth]{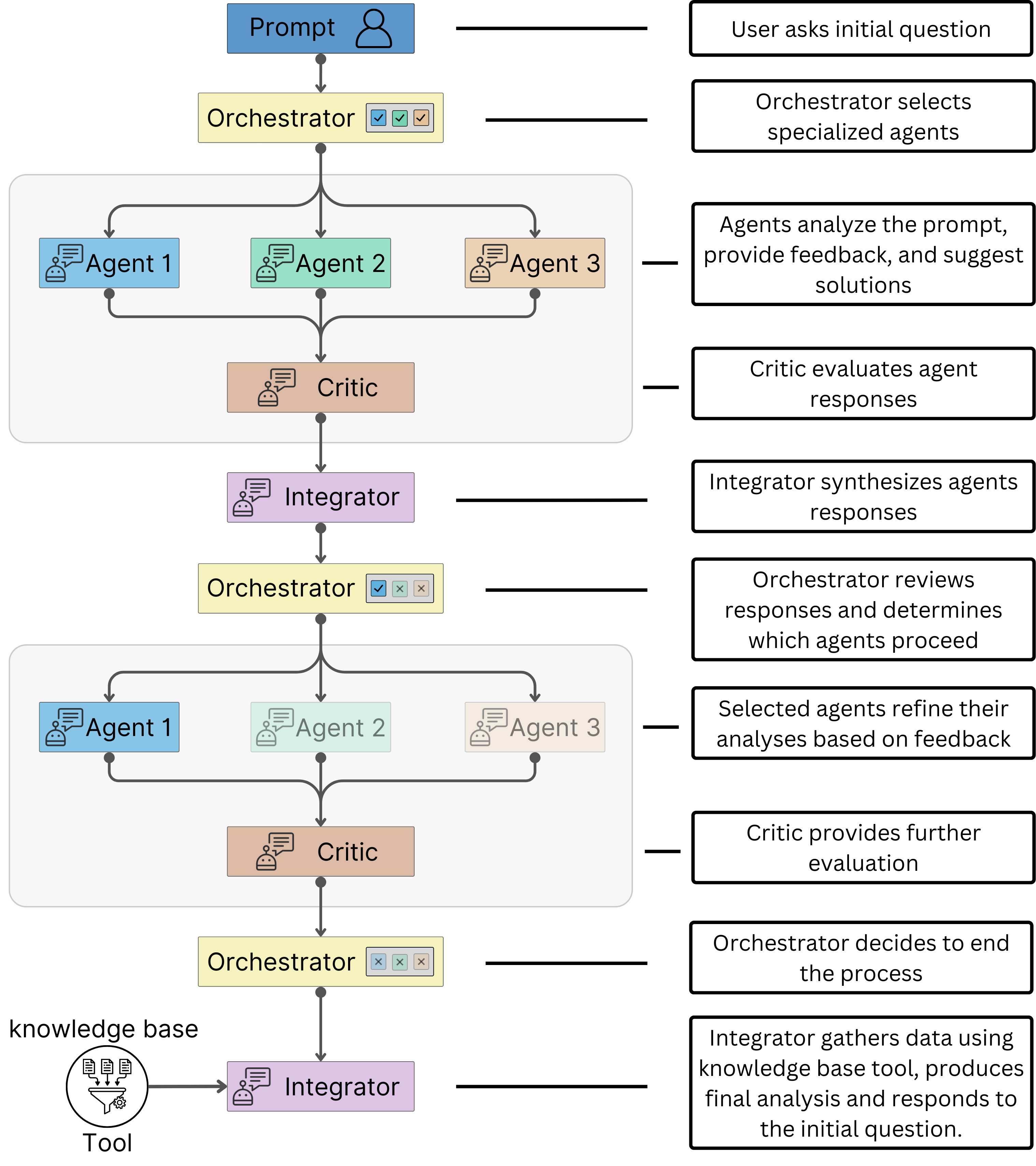}
\caption{System architecture showing the interaction among agents}
\label{fig:system_architecture}
\end{figure}

\begin{table*}[t]
\centering
\caption{Performance Metrics Across System Configurations}
\label{tab:results}
\begin{tabular}{l@{\hspace{1em}}c@{\hspace{1em}}c@{\hspace{1em}}c@{\hspace{1em}}c}
\toprule
\textbf{System Configuration} & \textbf{Argument Quality} & \textbf{Critical Engagement} & \textbf{Reference Cohesion} & \textbf{Risk Resolution (\%)} \\
\midrule
Base System & 2.75 & 0.00 & 1.13 & 0.00 \\
ToM Only & 3.15 & 0.21 & 2.63 & 22.50 \\
Critic Only & 3.00 & 0.21 & 2.38 & 26.25 \\
ToM + Critic & 3.43 & 0.38 & 2.50 & 48.75 \\
\bottomrule
\end{tabular}
\end{table*}

\subsection{Agent Roles and Decision-Making Process}

The system comprises three independent agents tasked with specific aspects of decision-making. The outputs they generate are stored in a systematically organized knowledge graph accessible only to the Integrator via the knowledge base tool.
Figure~\ref{fig:system_architecture} illustrates the architecture of our multi-agent system with the main features of the specialized agents, the Critic Agent, the Integrator, the Orchestrator and the external tools.

The \textbf{Data and Logic Specialist} analyzes the technical viability of investment alternatives. This includes dividing problems into measurable technical aspects, examining interdependencies among systems, and examining scalability limitations. Its argumentation is evidence-based, discovering the potential of implementation risks and ensuring that technological claims are relevant. When ToM is present, this agent anticipates the Visionary Strategist's concern with disruptive potential and the Implementation Realist's concern with feasibility, framing its argumentation to address their likely concerns.

The \textbf{Visionary Strategist} considers the long-term implications and market potential of the suggested technologies. Maps new use, industry transformation, and adoption patterns. In contrast to the Data and Logic Specialist, whose case rests on technical limitations, the Visionary Strategist concentrates on changing potential and competitive position strategy. When ToM is engaged, it expects the Data and Logic Specialist to be excessively cautious in assessing feasibility and the Implementation Realist to place a high priority on operational constraints. This allows it to frame its arguments in such a way as to accommodate these while maintaining its long-term orientation. 

The \textbf{Implementation Realist} focuses on resource availability, operational bottlenecks, and costs. Provides a pragmatic evaluation of investment feasibility, weighing the potential return on investment in light of the near-term risk. It challenges whether the proposed technologies are aligned with existing infrastructure and labor force capability. Where ToM is involved, it acts as a counterbalance to the Visionary Strategist's tendency to emphasize future opportunity and downplay current risks by introducing more realistic issues into the discussion. It also anticipates the Data and Logic Specialist to be more technically sound but possibly cavalier regarding practical deployment issues. 

The \textbf{Critic Agent} criticizes the responses of the specialized agents following the first round of reasoning, pointing out weaknesses, logical fallacies, and unsubstantiated statements. Unlike specialized agents, it does not advance its own solutions, but rather criticizes the prevailing arguments, inviting further improvement. Its role is analogous to peer review processes in human decision-making, in which criticism enriches reasoning by forcing individuals to re-examine assumptions. When ToM and structured critique are both active, the Critic Agent also evaluates the accuracy of agents' predictions regarding one another, identifying mismatches that can reveal flawed assumptions or coordination failures.

\subsection{Iterative Coordination and Decision Refinement}

The process utilizes an iterative mechanism of reasoning performed by two principal agents: Integrator and orchestrator.

The \textbf{Integrator} aggregates contribution of an agent into a coherent answer, with coherence ensured through organized retrieval and fulfillment of logical constraints. Unlike expert agents, Integrator can access an external tool consisting of a combination of GraphRAG and Clingo, combining retrieval over a knowledge graph with logical inference. After a response for each agent, relevant information is added to a Neo4j-based store, with a structured discussion model developed over a period of time by the system. In retrieval, the tool first transforms an agent query into a Cypher query to retrieve relevant information in a graph. In case information retrieval is lacking, a query is translated into an Answer Set Programming (ASP) format and run in Clingo, with logical inference performed for the retrieval of conclusions. In synthesis, balancing diversity and maintaining cohesion in an argument takes high priority. ToM performance is also monitored, with discrepancies between predicted and actual output detected between the output of agents and the actual output.

The \textbf{Orchestrator} determines whether refinement is necessary. In case of gaps in Integrator integration, it identifies the additional contribution of agents according to predefined heuristics. For instance, if technical feasibility of a proposed answer is doubtful, reactivation of the Data and Logic Specialist is initiated. In case uncertainty in the marketplace persists, reactivation of the Visionary Strategist is initiated. In addition, feedback received through the feedback of the Critic Agent is considered, with discrepancies in integration detected beforehand addressed in case a recommendation is generated.

The iterative process repeats until such a point when the discussion reaches a satisfactory level of resolution, according to the determination of the Orchestrator. Once all gaps have been addressed, the integrator provides a concluding synthesis, including key factors in deciding, comparing technology options, and providing an implementation roadmap.

\section{Experiments and Results}

To evaluate the contribution of ToM and Critic Agent to multi-agent collaborative reasoning, a series of experiments addressed a case study of a strategic technology investment decision for a hypothetical technology corporation with a \$12 million annual budget for investments in three emerging technology alternatives: Edge Computing, Quantum Computing, and Blockchain. Each technology involved specific technical, market and financial factors, with a balanced consideration of feasibility, potential impact, and concomitant risk required.

The model was run in four alternative settings: (FF) no ToM, no Critic, (TF) ToM alone, (FT) Critic alone, and (TT) both ToM and Critic. Evaluating the nuanced qualities of collaborative reasoning dialogues, such as argument structure, critical engagement, and cohesive integration of perspectives, presents challenges for traditional automated metrics. To enable a consistent and scalable assessment across the different system configurations, an external LLM-based judge rated the dialogues based on predefined requirements, unaware of the specific configuration being evaluated. To ensure consistency, the LLM judge was strictly limited to the provided rubric definitions and conversation content, explicitly instructed to avoid external knowledge lookups for its assessment. This approach allows for the application of a complex, qualitative rubric at scale, which would be resource-prohibitive with human evaluators alone

\subsection{Evaluation}

Conversations were rated with a rubric to elicit key dimensions of multi-agent reasoning. The LLM judge was tasked with assessing these qualitative dimensions: Argument quality was rated 1–5, with higher values for logically organized, supported, and persuasive reasoning. In addition to rating internal consistency, care was taken for responsiveness to prior turns and counterargument discussion. Engagement with critical thinking was measured in terms of the proportion of substantive counterarguments out of total turns, such that shallow criticism did not positively contribute to this metric. The cohesion in the references was rated 1–3, with a rating for how well the agents reflected and developed compared to each other, with larger values for a developed and sophisticated discussion. Risk resolution was measured in terms of the proportion of discovered risks addressed in a meaningful way in the concluding recommendation, and gauged the agent's ability to see and counter potential obstacles.

In addition to these numerical ratings, triggers for revision, in terms of an agent revising its position in response to a counterargument or a new constraint, were tracked. These triggers provided a glimpse into adaptability and iterative refinement capabilities of the model, and in terms of how much thinking changed in reaction to peer feedback. The identification of these events was performed by the LLM judge, so that any revisions were substantive and not shallow rewordings.

By comparing system performance under these contrasting scenarios, experiments hoped to tease apart ToM and structured critique contribution to multi-agent reasoning. The results form a basis for the evaluation of how simulation of the peer view and incorporation of critical evaluation inform group decision making in AI-powered groups.

\begin{table}[t]
\centering
\caption{Average Revision Triggers per Conversation}
\label{tab:behavior}
\begin{tabular}{l@{\hspace{1.5em}}c}
\toprule
\textbf{System Configuration} & \textbf{Revision Triggers} \\
\midrule
Base System & 0.00 \\
ToM Only & 1.00 \\
Critic Only & 0.50 \\
ToM + Critic & 1.75 \\
\bottomrule
\end{tabular}
\end{table}

\subsection{Results}

The TT configuration produced responses that, according to the LLM judge, rated the highest, thus confirming the synergistic value added from the engagement of ToM with a Critic Agent. It performed best, as depicted in Table~\ref{tab:results}, across criteria and brought about arguably the largest enhancements in argumentation and resolution of risks. In comparison, the bottom-performing baseline configuration, FF, where both variables were eliminated, had agents that failed to challenge one another's claims or to fuse different perspectives. TF and FT showed intermediate performance, suggesting that while ToM helped with coordination, the Critic Agent helped to sharpen arguments and identify undetected risks. Notably, the patterns found in the LLM judge's ratings—such as robust, iterative dialogues in configurations with critique (TT, FT) versus shallower, less adaptive dialogues without it (FF)—align with qualitative differences anticipated by cognitive theories of effective collaboration. 

Configurations without a Critic Agent (TF and FF) took fewer dialogue turns, suggesting faster convergence towards the final answer. This, however, indicates a fast-moving dialogue, but the FF configuration still, at the end of it, provided little depth of reasoning; dialogue brevity does not necessarily equal efficiency. Configurations with a Critic Agent (FT and TT) provided good feedback integration, but the best critique integration went for TT. 

Risk resolution was most affected by the different configurations. TT was able to show its best level of risk resolution with its strong capability to create robust strategies for handling uncertainties. In contrast, the FF system failed to resolve any risks but added other insights to the discussion without going into better-devised approaches. This stresses how much systematic criticism and predictive modeling enhance decision effectiveness. Table~\ref{tab:behavior} showed that the use of revision triggers had a very similar pattern, with TT showing the most adaptations for self-correction. ToM played a role in anticipating critique and proactively amending claims, with the Critic Agent being critical in resolving missed risks and contradictions aptly.

The findings demonstrate that ToM and structured critique together produce the most effective, well-supported, and critically evaluated recommendations. While ToM improved inter-agent alignment, the Critic Agent strengthened argumentation depth and risk assessment.

\begin{figure}[t]
\centering
\includegraphics[width=0.9\linewidth]{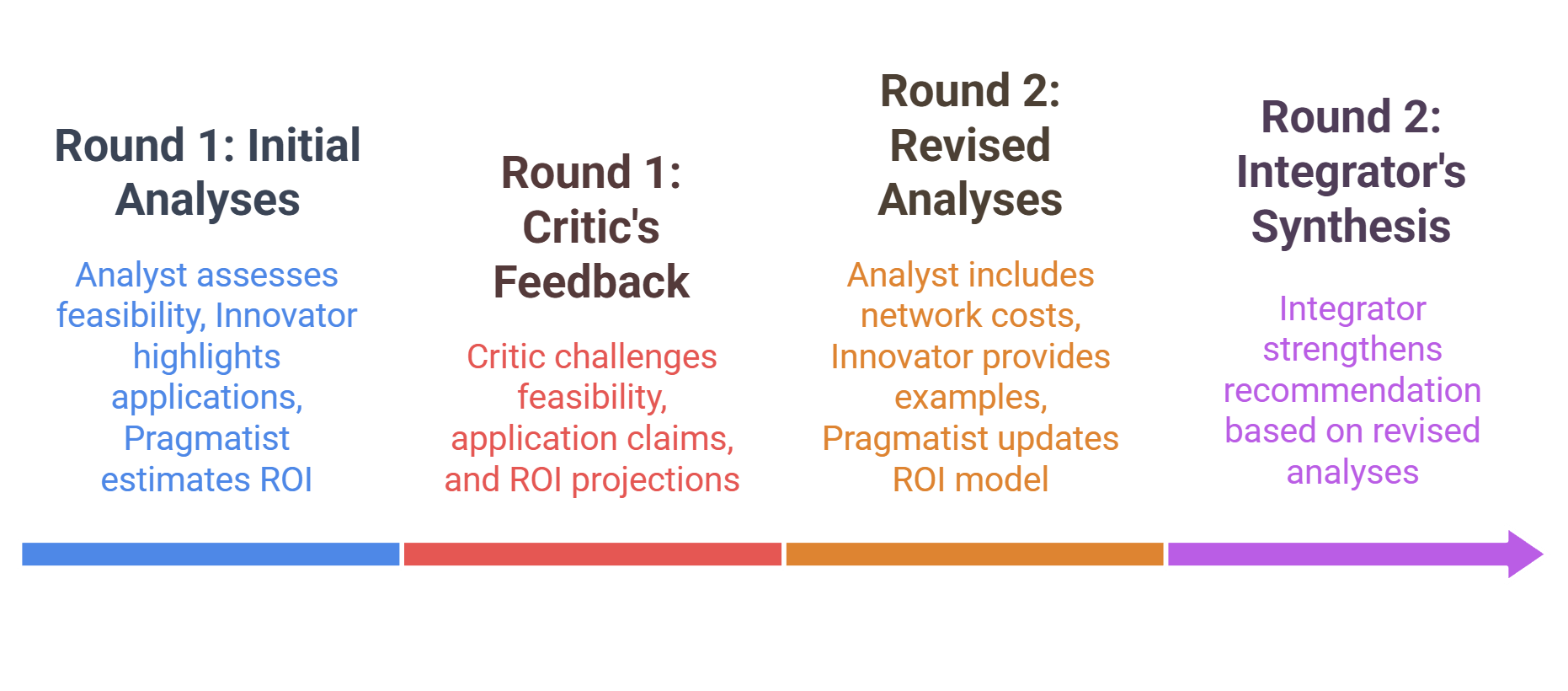}
\caption{Multi-Round Analysis Process with Critic-Driven Refinement}
\label{fig:coordination}
\end{figure}

\section{Discussion}

The experimental results reveal that both Theory of Mind (ToM) and the Critic Agent contribute to improving multi-agent reasoning, but through different processes. The best configuration (TT) merged both modules, producing more holistic output, showing increased risk consideration and higher adaptability. Notably, neither module alone guaranteed high performance, leading to variable results in configurations testing only a single module (ToM-only and Critic-only).

The integration of ToM significantly improved efficiency in coordination and referential cohesion. The agents forming a part of ToM configurations (TF and TT) produced more relevant references to previous statements, which indicated that ToM promotes better coordination and reduced unnecessary exchanges. But then again, ToM could not guarantee high-quality output in terms of producing high-quality arguments and effective resolution of risks, meaning that even with better coordination, thorough critical thinking requires systemic critique.

The Critic Agent was key for revisions and argument quality improvement (Fig \ref{fig:coordination}), leading to deeper analysis and stronger conclusions. Critic configurations (FT, TT) showed better feedback integration and risk resolution. Integration depth was greatest in TT, suggesting that ToM might even enhance processing of critiques, consistent with theories relating perspective-taking~\cite{WoolleyEvidenceFA} to the integration of diverse feedback.

Despite improvements, the challenge with risk resolution remained (max 48.75\% in TT), pointing to limitations in managing uncertainties even with ToM and Critique. The baseline FF configuration was notably unreliable in risk resolution, underscoring the need for verification. These results were derived from a single case study, hence they need further exploration regarding generalizability across domains.

Due to the brevity of the interaction, the influence of the knowledge base tool in this study was small. Utilized in retrieving logically coherent conversation data, it was not utilized to its full potential in the Integrator. In long-term interaction situations with much higher requirements in retrieving information, the tool could contribute even more by extracting specific references and strengthening the checking of logical coherence. Future scenarios requiring iterative refinement and long-term recall should explore its impact.

A key challenge experienced was the premature closing of discussions in the absence of a Critic Agent. In both TF and FF, the Orchestrator Agent consistently interrupted the dialogue before all critical aspects were discussed, most likely due to the lack of structured critique encouraging continuous refinement. Since its role is to assess the adequacy of responses, its decision-making basis can be optimized, particularly in settings that lack structured critique. Further research could explore adaptive routing techniques, ensuring critical issues have been discussed before interaction termination.

%end

\section{Conclusion and Future Work}

A multi-agent reasoning schema that incorporates ToM and structured critique to support joint decision-making is presented in this study. The findings indicate that the proposed mechanisms enhance coordination, argument quality, and risk analysis, ultimately leading to more well-structured and justifiable solutions. The coordination mechanisms used in this study are considerably looser compared to the traditional multi-agent systems that rely on rigid forms of coordination~\cite{li-etal-2023-theory, lee2024unifieddebuggingapproachllmbased}. ToM maximizes alignment and minimizes redundancy, while the Critic Agent enhances reasoning depth by prompting adjustments and highlighting overlooked risks. Their combined effect produced the strongest responses, thus demonstrating that integrating predictive modeling with structured critique is a major contributor to success.

While this is good news, some limitations need to be pointed out. First, the study is based on one case analysis; thus, future studies should be directed to increase the generalizability of the results to an array of contexts. Second, using an LLM-based judge carries with it a recognized set of limitations~\cite{10.5555/3666122.3668142, gu2025surveyllmasajudge}, and the results even lacked direct validation against expert human annotation for this specific task; hence, human validation should become a priority for future work. Third, although cognitive mechanisms have improved interaction quality, the scalability of the framework to significantly larger and more complex multiagent settings remains a matter of inquiry that needs dedicated attention. Future research should thus validate the framework across different domains and scales, with defined evaluation metrics and statistical analyses, for widespread applicability.

Theoretically, beyond multiagent reasoning, this framework has broader potential for adaptive AI-supported decision-making, such as autonomous negotiation, policy analyses, and advisory system builds wherein agents must consider myriad perspectives while holding each perspective to rigor. Here, the capacity to dynamically anticipate contradicting peer reasoning while engaging in structured criticism can also aid domains like legal reasoning, collaborative research, and education~\cite{ChudziakKostkaAIED2025}, wherein decision making is often an iterative round of refining and assaying risk.

A future line of work aims toward the refinement of ToM mechanisms to model reasoning styles and strategical preferences more accurately. The improvement of the orchestrator agent's decision-making procedures could facilitate further improvements in conversation and negotiation flow, ensuring discussions arrive at well-supported conclusions without premature termination. In addition, this framework should be validated in real-world settings involving multiple agents across software engineering, legal reasoning, and education to establish its impact~\cite{10.1145/3691620.3695336}.

This study advances toward developing AI systems that are capable of deeper forms of human-like collaborative reasoning. The cognitive synergy explored here opens new horizons for developing an AI that is truly transparent, flexible, and efficient in resolving complex decision-making tasks.

%end

\section{Acknowledgments}
The work reported in this paper was partially supported by the Polish National Science Centre under grant 2020/39/I/HS1/02861.

\bibliographystyle{apacite}

\setlength{\bibleftmargin}{.125in}
\setlength{\bibindent}{-\bibleftmargin}

\bibliography{CogSci_Template}

\end{document}